\documentclass[runningheads,a4paper]{llncs}

\usepackage{amssymb}
\usepackage{graphicx}

\usepackage{url}
\urldef{\mailsa}\path|{yang.wang, jasmin.ramadani, stefan.wagner}@informatik.uni-stuttgart.de|    
\newcommand{\keywords}[1]{\par\addvspace\baselineskip
\noindent\keywordname\enspace\ignorespaces#1}
\usepackage{booktabs} 
\newcommand{\tabincell}[2]{\begin{tabular}{@{}#1@{}}#2\end{tabular}}
\usepackage{multirow}
\usepackage{slashbox}

\begin{document}

\mainmatter  

\title{An Exploratory Study on Applying a Scrum Development Process for Safety-Critical Systems}
\titlerunning{An Exploratory Study on Applying Scrum for Safety-Critical Systems}

%
%
\author{Yang Wang
\and Jasmin Ramadani\and Stefan Wagner}
\authorrunning{}

\institute{University of Stuttgart, Germany
\mailsa\\
}

%
%

\toctitle{}
\tocauthor{}
\maketitle

\begin{abstract}
\small
\emph{Background:} Agile techniques recently have received attention from the developers of safety-critical systems. However, a lack of empirical knowledge of performing safety assurance techniques, especially safety analysis in a real agile project hampers further steps. 
\emph{Aims:} In this article, we aim at (1) understanding and optimizing the S-Scrum development process, a Scrum extension with the integration of a systems theory based safety analysis technique, STPA (System-Theoretic Process Analysis), for safety-critical systems; (2) validating the Optimized S-Scrum development process further. 
\emph{Method:} We conducted a two-stage exploratory case study in a student project at the University of Stuttgart, Germany.   
\emph{Results:} The results in stage 1 showed that S-Scrum helps to ensure safety of each release but is less agile than the normal Scrum. We explored six challenges on: priority management; communication; time pressure on determining safety requirements; safety planning; time to perform upfront planning; and safety requirements' acceptance criteria. During stage 2, the safety and agility have been improved after the optimizations, including an internal and an external safety expert; pre-planning meeting; regular safety meeting; an agile safety plan; and improved safety epics and safety stories. We have also gained valuable suggestions from industry, but the generalization problem due to the specific context is still unsolved.

\end{abstract}

\vspace{-0.6cm}
\section{Introduction}
\vspace{-0.2cm}
To reduce the risks and costs for reworking and rescheduling, agile techniques have aroused attention for the development of safety-critical systems. Traditionally standardised safety assurance, such as IEC 61508 \cite{iec61508}, is based on the V-model. Even though there is no prohibition to adapt standards for lightweight development processes with iterations, some limitations cannot be avoided during the adaptation \cite{turk2014limitations}. 
Existing research in agile techniques for safety-critical systems is striving for consistency to standards. Safe Scrum \cite{staalhane2012application} is a considerable success due to a comprehensive combination between Scrum and IEC 61508. However, an integrated safety analysis to face the changing architectures inside each sprint still needs to be enhanced. Therefore, in 2016, we proposed S-Scrum to integrate a systems theory based safety analysis technique, STPA (System-Theoretic Process Analysis) \cite{leveson2011engineering}, which was proposed by Leveson in 2012, inside each sprint to guide a safe design \cite{wang2016toward}.  \newline \newline
\emph{Problem statement.} We proposed to integrate STPA in a Scrum development process to enhance the safety in agile development. However, it has not been validated in practice. As far as we know, there exists no empirical data on applying Scrum for a safety-critical project with the integration of STPA. \newline   \newline   
\emph{Research objective and research questions.} In this article, we aim to explore the agility and safety of S-Scrum as well as challenges and their relevant optimizations for developing a safety-critical system called ``Smart Home". The research questions are as follows: \newline 
\textbf{RQ 1} \emph{How does S-Scrum handle agility and safety in safety-critical systems?}\newline 
\textbf{RQ 2} \emph{What are the challenges of S-Scrum in such a context?}\newline  
\textbf{RQ 3} \emph{How could S-Scrum be optimized to overcome the challenges?}\newline
\textbf{RQ 4} \emph{What are the effects of the optimized S-Scrum on safety and agility?}\newline \newline
\emph{Contribution.} This paper provides the first case study on applying a Scrum development process for safety-critical systems. We investigated the effects and challenges of S-Scrum in the 1st stage of the case study. We proposed an optimized S-Scrum and validated it in the 2nd stage of the case study. To this end, we preliminarily discussed the optimized S-Scrum in industry.\newline\newline
\emph{Outline.} The paper is organized as follows. First, we present the related work on using Scrum for safety-critical systems and normal Scrum development process improvement (Sect. 2). Then, we present the background about STPA and our previous work about S-Scrum (Sect. 3). After that, we describe the approach and results of the 1st stage of the case study (Sect. 4.1), and the 2nd stage of the case study (Sect. 4.2). Finally, we discuss the threats to validity (Sect. 5), and draw the conclusions (Sect. 6).

\vspace{-0.2cm}
\section{Related Work}
\vspace{-0.2cm}

To the best of our knowledge, few empirical studies of applying Scrum or other agile processes for safety-critical systems exist. Most of the research is still in the stage of theoretical illustration and validation \cite{ge2010iterative} \cite{vuori2011agile}. \par
Safe Scrum is a Scrum development process for safety-critical systems, which was developed to adhere to the general functional safety standard IEC 61508 \cite{staalhane2012application} \cite{hanssen2016quality}. Previous research of Safe Scrum has been synergized with other safety standards in different domains \cite{staalhane2013scrum} \cite{staalhanesafety}. However, purely theoretical validation is unable to cover the details of the process. More practical experiences are becoming crucial.  \par
Despite the limited practical experiences in applying Scrum for safety-critical systems, there are a lot of Scrum development process experiences that could be taken as a reference for the agile software process improvement of our project \cite{Moe:2010:TMU:1752257.1752480} \cite{begel2007usage}. Diebold et al. \cite{diebold2015practitioners} investigated the industrial usage of Scrum with various sprint length, events, team size, requirements engineering, roles, effort estimations and quality assurance. Cho \cite{cho2010exploratory} conducted an in-depth case study in two organizations. The data was analyzed along 4 dimensions, including human resource management; structured development process; environment; information systems and technology. These factors were covered in our assessment of agility considering our criteria to improve the S-Scrum. \par

\vspace{-0.2cm}
\section{STPA and S-Scrum}
\vspace{-0.2cm}

STPA is a new hazard analysis technique by Leveson in 2012. It has been successfully used in various domains, such as aviation, automobiles and healthcare. Compared with the traditional safety analysis techniques, such as FMEA (Failure Mode and Effects Analysis) and FTA (Fault tree analysis), STPA bases on the systems theory rather than the traditional reliability theory. Due to an increasing complexity of systems, the accidents are not caused by single function failures or chains of failure events, but resulted from inadequate control actions. To ensure the safety of today's complex systems, the use of STPA is becoming necessary. Besides, we proposed using STPA in a Scrum development process \cite{wang2016toward}, as current safety analysis techniques start from a complete design, which is not consistent to agile methodologies, which advocate a lightweight up-front planning and design. STPA, on the contrary, provides the necessary information to start from a high-level architecture and to guide the incremental design process. In S-Scrum, we integrate STPA mainly in three aspects: (1) During each sprint, we integrate STPA as safety-guided design. (2) At the end of each sprint, we use STPA on the product instead of a Reliability, Availibility, Maintainability and Safety (RAMS) validation. (3) We replace the final RAMS validation with STPA. The other parts are kept consistent to Safe Scrum: (1) The environment description and the SSRS phases 1-4 (concept, overall scope definitions, hazard and risk analysis and overall safety requirements). (2) Test Driven Development. (3) Safety product backlog. (4) A safety expert \cite{wang}. We aim to fill the gap of a lack of safety analysis in agile development and enhance the safety on the basis of a standard-based Scrum development process for safety-critical systems.     

\vspace{-0.2cm}
\section{Case Study}
\vspace{-0.2cm}

To explore S-Scrum further, we conduct this study following the guideline by Runeson \cite{runeson2009guidelines} and Yin \cite{yin2013case}. We design this case study with a multi-staged procedure. Each stage has different objectives and research questions. We explored the challenges and optimizations in \textbf{S-Scrum} in stage 1, while we validated the \textbf{optimized S-Scrum} in stage 2.

\vspace{-0.2cm}
\subsection{Research Context}

The case study (including stage 1 and stage 2) was performed in the project developing safety-critical systems, Smart Home, between March, 2016 and March, 2017 at the Institute of Software Technology, University of Stuttgart. The project had 400 planned working hours
per head with a headcount of 14 students. The students have taken part in a training program for agile development and STPA before joining the project and a course on automation systems during the project. The Scrum Master was one research assistant with experienced project management background, while the Product Owner and Safety Expert was another research assistant majoring in using agile for safety-critical systems. All the students were supervised by three research assistants. The project was to work on an IoT based smart home with a smart coffee machine, smart light alarm system, autonomous parking system, door-open system, and smoke detector alarm system through the IoT server - KAA\footnote{https://www.kaaproject.org/overview/}. The project ``Smart Home" is openly available in GitHub\footnote{https://github.com/ywISTE/student-project---Smart-Home}. 

\vspace{-0.2cm}
\subsection{Case study - stage 1}

The objective of stage 1 is to validate the safety and agility of S-Scrum and optimize it. In stage 1, we focus on answering RQ 1, RQ 2, and RQ 3. The general research strategy in stage 1 is shown in Table 1. 

\begin{table}[!h]
\scriptsize
\center
\caption{Research strategy in stage 1 (``DL"-Developer, ``SH"-Stakeholder, ``SM"-Scrum Master)}
\begin{tabular}{l|l|l|l|l}

  \toprule
  \textbf{Time}& Sprint 1 to sprint 5 & Sprint 6 to sprint 7 & Sprint 8 & Sprint 9\\ \toprule
  \textbf{Process}    & Scrum & S-Scrum & S-Scrum & S-Scrum \\ \hline  
  \textbf{\tabincell{l} {Data\\ collection}} & \tabincell{l} {Participant observation \\ Scrum artifacts \\ Documentation review}& \tabincell{l} {Participant observation \\ Scrum artifacts \\ Documentation review} & \tabincell{l} {Questionnaires} & \tabincell{l} {Semi-structured \\ interviews} \\ \hline
  \textbf{Participants} & \tabincell{l} {DLs \\ SHs}& \tabincell{l} {DLs \\ SHs} & \tabincell{l} {13 voluntary DLs} & \tabincell{l} {5 voluntary DLs \\ 1 SM} \\ \hline
  \textbf{\tabincell{l} {Data\\ types}}    & Quantitative & Quantitative & Quantitative & Qualitative \\ \hline 
  \textbf{Analysis}    & Sum of the numbers & Sum of the numbers & \tabincell{l} {Median \\MAD} & Coding \\ \hline   
  \textbf{Output} & \tabincell{l} {No safety data}& \tabincell{l} {Safety data: \\ M16.1-M16.3 \\M17.1-M17.3} & \tabincell{l} {Agility data: \\M1-M15} & \tabincell{l} {Challenges \\ and \\ optimizations \\of S-Scrum} \\ 
\bottomrule
\end{tabular}
\end{table}%

\vspace{-0.4cm}
\subsubsection{Data collection in stage 1}
\vspace{-0.2cm}

Stage 1 spans from sprint 1 to sprint 9. Each sprint lasts three weeks. The agility-related quantitative data, M1 to M15, were collected through 13 questionnaires\footnote{The questionnaire is available: https://zenodo.org/record/439696\#.WODCovl96Uk}. Our participant observation as the Product Owner (the first author), the Scrum Master, and the customer imposed also an evaluation and review of the results. The safety-related data, M16.1 to M16.3 and M17.1 to M17.3, were quantitatively collected during sprint 6 and sprint 7. From sprint 1 to sprint 5, we executed normal Scrum without safety analysis for the adaptation and preparation for the project. The STPA was performed by the safety expert and recorded privately by using the STPA tool, XSTAMPP\footnote{http://www.xstampp.de/}, while the hazards and safety requirements were recorded in the safety product backlog in Jira. \par 
Based on the quantitative data for agility and safety, we then designed semi-structured interviews with 6 voluntary participants from the development team, including the Scrum Master and five developers. The interviews lasted 270 minutes overall. The questions began with a specific set of questions regarding the observations. Further, we asked about the causalities. Finally, the optimizations were collected in an open-ended mode. The interview guideline\footnote{The interview guideline is available: https://zenodo.org/record/439696\#.WODCovl96Uk} was provided before each interview. We recorded interview data in field notes and we used the audio recordings for text transcription. 
\vspace{-0.2cm}
\subsubsection{Data analysis in stage 1}
\vspace{-0.2cm}

We analyzed the data using the combination of GSN \cite{kelly2004goal} and GQM \cite{basili1992software} referring partially to the VMF framework \cite{cruickshank2009validation}, as shown in Fig. 1. The data are from two aspects: agility (S1) and safety (S2). To evaluate and optimize agility (S1), we set 15 goals (G1 to G15) considering Comparative Agility Survey \cite{williams2010driving}. They are: G1 (Team work composition); G2 (Team work management); G3 (Communication); G4 (Requirement emergency); G5 (Technical design); G6 (Planning levels); G7 (Critical variables); G8 (Progress tracking); G9 (Sources of dates and estimates); G10 (When do we plan); G11 (Customer acceptance test); G12 (Timing); G13 (Quality focus); G14 (Reflection); G15 (Outcome measure). To reach G1 to G15, we analyzed M1 to M15 indirectly by setting sub-metrics. For example, M1 (Team work composition) was analyzed by M1.1 (Team members are kept as long as possible), M1.2 (Specialists are willing to work outside their specialty to achieve team goals), M1.3 (Everyone required to go from requirements to finished system is on the team), and M1.4 (People are no more than two teams). Each sub-metric was analyzed on an ordinal scale of 5 (e.g., from 1 to 5 means ``Negative", ``More negative than positive", ``Neither negative nor positive", ``More positive than negative", and ``Positive"). To investigate the in-depth challenges, we found out either the negative values of the results or the significant differences between the normal Scrum and S-Scrum to formulate further interview questions. To analyze the interview results, we used NVivo11 for text encoding \cite{strauss1997grounded}. 
Concerning safety, G16 is extended with 3 questions together with 3 metrics including: number of software hazards (M16.1), number of software safety requirements (M16.2), and number of safety requirements traceable to hazards (M16.3). G17 is extended to be evaluated by the number of mitigated hazards (M17.1), number of accepted safety requirements (M17.2) in the present sprint, and number of rejected safety requirements (M17.3) in the project. 

\begin{figure*}[!hbt]
\center
\includegraphics[width=0.8\textwidth]{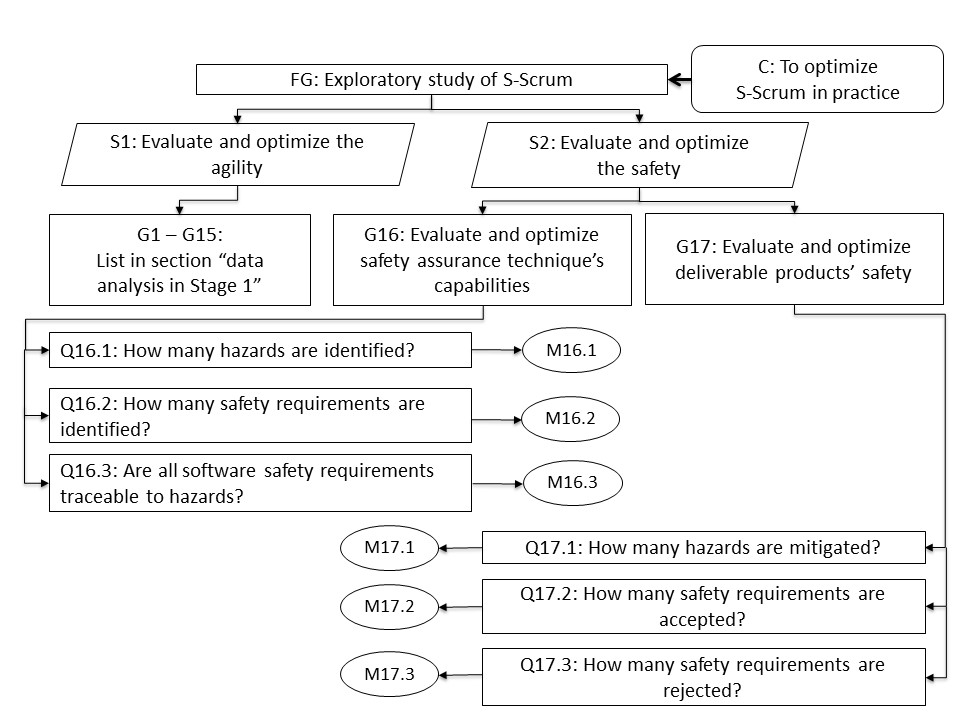}
\caption{General data analysis strategy (``FG"-Final Goal, ``S"-Strategy, ``G"-Goal, ``C"-Context, ``Q"-Question, ``M"-Metric )}
\end{figure*}

\vspace{-0.2cm}
\subsubsection{Results in stage 1 - RQ 1: How does S-Scrum handle agility and safety in safety-critical systems?  }

We investigate the effect on agility by comparing the normal Scrum and the S-Scrum according to the 15 metrics in Fig. 2. From the general overview, we can conclude that most of the values regarding agility in S-Scrum are slightly worse than those in the normal Scrum, while one metric shows strongly negative values (``when do we plan"). We discussed the results with the technical support from the Comparative Agility Survey and got the feedback: \emph{when most of the values are more positive than negative (more than ``3"), we could say that the process is agile enough.} Moreover, most values show relatively small differences between normal Scrum and S-Scrum. Thus, we consider the agility of S-Scrum to be acceptable. Yet, optimizations are needed.       
Regarding the safety of S-Scrum, we performed STPA two rounds in sprint 6. We found 6 software hazards (M16.1) and 15 safety requirements (M16.2), which can all be traced back to software hazards (M16.3). Three hazards were mitigated (M17.1), while 14 safety requirements were accepted (M17.2). In sprint 7, we performed two rounds of STPA analysis. We found 10 software hazards (M16.1) and 24 safety requirements (M16.2), which can also all be traced back to software hazards (M16.3). Six hazards were mitigated (M17.1), while 23 safety requirements were accepted (M17.2). Each sprint has 1 rejected safety requirement due to hardware limitation (M17.3). \par 
\begin{figure*}[!hbtp]
\center
\includegraphics[width=1.0\textwidth]{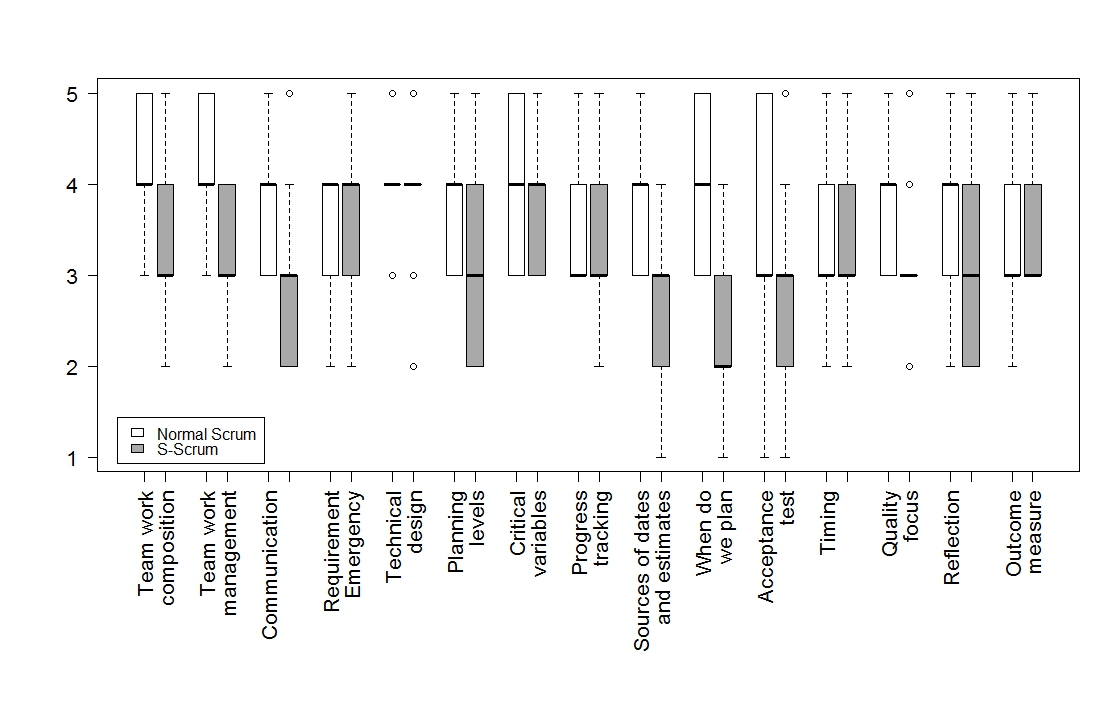}
\caption{Boxplots for general agility comparison between normal Scrum and S-Scrum (From ``1" to ``5" means less agile (``negative") to very agile (``positive"))}
\end{figure*}

\vspace{-0.2cm} 
\subsubsection{Results in stage 1 - RQ 2 \& RQ 3: What are the challenges of S-Scrum in such context? \& How could S-Scrum be optimized to overcome the challenges? }
  
To optimize S-Scrum, we derived six challenges from the six abnormal values (see data analysis in stage 1) from the sub-metrics inside these 15 metrics.  \newline  
\emph{Challenge 1: The priority management of safety requirements and functional requirements has conflict.} 
In the normal Scrum, the management and development team determine the sprint backlog with functional requirements in the sprint planning meeting. All the team members have a clear overview of and commitment to the sprint plan with relatively high-level features. The developers accomplish each item with their own detailed tasks. The requirements from the management and the concrete realizations from the developer reach a consensus during each sprint. In S-Scrum, the integrated STPA and the safety requirements break the balance. The functional requirements are correlated with the safety requirements. However, some developers preferred: \emph{functional requirements are more important than the safety requirements.} It was found that the need for long-term quality was given a lower priority than the need for short-term progress \cite{moe2012challenges}. Moreover, the safety expert spent a relatively short time working with the team members which influences also the decision making. As one developer mentioned: \emph{The safety expert is not working in the same room with the development team and has an inconsistent working time.} Thus, a lack of an in-time decision maker on the safety requirements together with the ignorance of safety requirements in the development team cause the conflict.\newline
To face this challenge, a \textbf{safety culture} should be integrated into a light-weight development process. We suggest to include an \textbf{internal safety expert} in the development team to (1) spread the safety culture; (2) increase the safety expert's working time with the team members; (3) clarify the bewilded safety requirements. An \textbf{external safety expert} is necessary to keep the communication with other stakeholders. To fill the gap between the external safety expert and the development team, the development team suggests that the external safety expert should join at least once the weekly Scrum meeting. The discussion between the management, the external safety expert and the internal safety expert could strive a fresh balance on the priorities. \newline      
\emph{Challenge 2: The communication between team members and safety expert is disturbed.}
To start with, the unclear safety-related documentation influences an effective communication. The team members mentioned: \emph{it is difficult to comprehend the purpose of the safety expert and integrate into our daily work from the existing documents.} Moreover, a lack of safety-related knowledge of the development team influences the discussion concerning safety issues. Finally, the insufficient time spent between safety expert and development team causes also a poor communication. Without a non-obstacle work place to communicate within the team about the work progress, the safety assurance could either be a superficial decoration or even worse, a roadblock during fast product delivery.\newline  
To face this challenge, in addition to the separated \textbf{internal safety expert} and \textbf{external safety expert}, a \textbf{weekly safety meeting} is suggested by an interviewee: \emph{The internal safety expert and external safety expert should meet each other at least once a week to exchange the status of the development team. Because the discussion should be deep in the safety area, it is not supposed to be established during the normal weekly Scrum meeting.} Last but not least, we improve our \textbf{safety epics} and \textbf{safety stories} to support an effective communication \cite{wangdc}, as shown in Sect. 4.3 (Optimized S-Scrum). \newline   
\emph{Challenge 3: The safety requirements are not determined early enough to appropriately influence design and testing.} 
In sprint 6 and sprint 7, the safety requirements were determined by the development team and the safety expert together in the sprint planning meeting. However, as one interviewee mentioned: \emph{the determination of safety requirements from the safety product backlog is too late to avoid a conflict between the functional requirements and their suitability for the coming sprint}. Thus, sometimes the functional design and testing have to start without the in-time safety requirements. \newline
To face this challenge, we propose \textbf{a pre-planning meeting} for solving the time pressure problem. First, the internal, external safety experts and product owner discuss the safety product backlog and the functional product backlog in the pre-planning meeting. Then they brainstorm the results with the whole development team in the sprint planning meeting to gather more ideas and make each safety requirement clear. \newline 
\emph{Challenge 4: The planning at the start of each iteration is insufficient.}
In the normal Scrum, the development team and the product owner plan the upcoming sprint in the sprint planning meeting by formulating the sprint backlog with estimated items, which makes the development team sufficiently confident about their plan. 
However, the estimation and planning for the safety product backlog seem not ideal, as well as the interconnection with the functional product backlog, which make an in-time identification of the sprint backlog difficult. An interviewee said: \emph{It is difficult to determine the safety requirements when the development team has not planned the functional requirements for the coming sprint.} \newline
To face this challenge, we suggest and adapt an \textbf{agile safety plan} \cite{myklebust2016agile} in connection with the \textbf{pre-planning meeting} to increase the understanding of safety issues and enhance confidence. In our project, the results of STPA are part of the agile safety plan. \newline 
\emph{Challenge 5: The time to perform upfront planning is late.}
A team member said: \emph{the pre-planning meeting for safety issues should start before the sprint planning meeting. But the concrete time should be decided between the external safety expert, the internal safety expert and the product owner.} Based on the experience of the previous sprints, it is better to start upfront planning one week before the sprint planning meeting (3 weeks/sprint). The time could be changed depending on the sprint length. More explanations are in challenge 4. \newline
\emph{Challenge 6: The safety requirements lack well-defined completion criteria.}
In the normal Scrum, we have various testing methods to determine the completion of each feature such as unit testing, system testing, regression testing, and acceptance testing, which are promoted to be automated in an agile context. However, few agile testing methods are suitable for validating safety requirements, as the safety requirements are either from standard requirements or the safety analysis, which differentiates safety testing and functional testing. In S-Scrum, we use UAT (User Acceptance Testing) for validating safety requirements. Thus, a suitable safety criterion becomes important.  \newline
To face this challenge, we use a ``Given-When-Then" format \cite{garg2015cucumber} as \textbf{safety requirements' criteria}. The development team suggest that the external safety expert could decide the \textbf{safety stories' criteria} and the internal safety expert could decide the \textbf{safety tasks' criteria}. The whole development team could \textbf{brainstorm} both criteria. To this end, the product owner and safety expert perform the acceptance testing. 
\vspace{-0.2cm}
\subsection{Case study - stage 2}

After the optimizations described above, the objective of stage 2 is to validate the safety and agility of the optimized S-Scrum and discuss it in industry. We focus on answering the RQ 4 together with some discussion from industry. The general research strategy in stage 2 is shown in Table 2.

\begin{table}[!hbt]
\scriptsize
\center
\caption{Research strategy in stage 2 (``DL"-Developer, ``SH"-Stakeholder, ``SM"-Scrum Master, ``PO"-Product Owner)}
\begin{tabular}{l|l|l|l}

  \toprule
  \textbf{Time}& Sprint 10 to sprint 11 & Sprint 12 & Sprint 13\\ \hline
  \textbf{Process}    & optimized S-Scrum & optimized S-Scrum & optimized S-Scrum \\ \hline  
  \textbf{Data collection} & \tabincell{l} {Participant observation \\ Scrum artifacts \\ Documentation review}& \tabincell{l} {Questionnaires} & \tabincell{l} {Semi-structured \\ interviews} \\ \hline
  \textbf{Participants} & \tabincell{l} {DLs \\ SHs}& \tabincell{l} {8 voluntary DLs} & \tabincell{l} {1 PO (from EPLAN) \\ 1 SM (from EPLAN)} \\ \hline
  \textbf{Data types}    & Quantitative & Quantitative & Qualitative \\ \hline 
  \textbf{Analysis}    & \tabincell{l} {Sum of the numbers \\ (compare with the data \\from stage 1)} & \tabincell{l} {Median and MAD \\ (compare with the data \\from stage 1)} & Coding \\ \hline   
  \textbf{Output} & \tabincell{l} {Safety data: \\ M16.1-M16.3 \\M17.1-M17.3} & \tabincell{l} {Agility data: \\M1-M15} & \tabincell{l} {Preliminary\\discussion \\ in industry} \\ 
\bottomrule
\end{tabular}
\end{table}%

\vspace{-0.2cm}
\subsubsection{Optimized S-Scrum}
\vspace{-0.2cm}

To have a clear overview, we compare the optimized S-Scrum to the normal Scrum and the S-Scrum in our project respectively in Table 3. 
In the optimized S-Scrum, we differentiate between an internal safety expert and an external safety expert. A pre-planning meeting and weekly safety meetings are established between safety experts. We include the safety epics, to satisfy $<$the overall safety needs$>$, the system must $<$always be able to reach a safe state$>$ \cite{myklebust2016safety}, in the story map. The safety product backlog is improved with optimized safety story: To keep $<$control action$>$ safe, the system must $<$achieve or avoid something$>$. An agile safety plan based on STPA technology is suggested for a clear overview. The safety culture is expected to be enhanced by the additional activities.   

\begin{table}[!h]
\scriptsize
\center
\caption{Normal Scrum, S-Scrum and optimized S-Scrum in Smart Home (``DL"-Developer, ``SM"-Scrum Master, ``PO"-Product Owner, ``SE"-Safety Expert)}
\begin{tabular}{l|l|l|l}

  \toprule
  \textbf{\tabincell{l} {Normal \\Scrum}}& \tabincell{l} {14 DLs\\1 SM\\1 PO} & \tabincell{l} {Sprint planning meeting\\Weekly Scrum meeting (2 times/week)\\Sprint review meeting\\Sprint retrospective meeting}& \tabincell{l} {Story map\\Product backlog\\Sprint backlog}\\ \hline
  \textbf{S-Scrum}   & \tabincell{l} {14 DLs\\1 SM\\1 PO\\1 SE} & \tabincell{l} {Sprint planning meeting \\(with safety planning)\\Weekly Scrum meeting (2 times/week) \\(with safety discussion)\\Sprint review meeting \\(with safety review)\\Sprint retrospective meeting}& \tabincell{l} {Story map\\Functional product backlog\\Safety product backlog\\Sprint backlog}\\ \hline
  \textbf{\tabincell{l} {Optimized \\S-Scrum}} & \tabincell{l} {13 DLs\\1 SM\\1 PO\\1 external SE\\1 internal SE}& \tabincell{l} {Pre-planning meeting \\Sprint planning meeting\\(brainstorming requirements and criteria)\\Weekly Scrum meeting (2 times/week) \\Weekly safety meeting (1 time/week)\\Sprint review meeting \\(with safety review)\\Sprint retrospective meeting}&\tabincell{l} {Story map\\(with safety epics)\\Functional product backlog\\Safety product backlog\\(with safety stories)\\Sprint backlog\\Safety plan} \\ 
  \bottomrule

\end{tabular}
\end{table}%

\vspace{-0.2cm}
\subsubsection{Data collection in stage 2}
\vspace{-0.2cm}

Stage 2 is from sprint 10 to sprint 13. The safety-related data, M16.1 to M16.3 and M17.1 to M17.3, were collected in the same way as in stage 1. The safety results were collected by both internal and external safety experts. The agility-related data, M1 to M15, were collected by the second round questionnaires\footnote{The questionnaire is available: https://zenodo.org/record/439696\#.WODCovl96Uk}. We further discussed the optimized S-Scrum by conducting 2 semi-structured interviews with one Scrum Master and one Product Owner from EPLAN GmbH, Germany. The interview lasted 2 hours. We formulated questions about the status of the Scrum development process in the company projects; the feasibility of the optimized S-Scrum in industry; and further suggestions from the industrial perspective. A project background illustration was provided before the interviews, together with the interview guidelines\footnote{The interview guideline is available: https://zenodo.org/record/439696\#.WODCovl96Uk}. The field notes, interview transcripts, and voice recordings were all preserved for backup. 

\vspace{-0.2cm}
\subsubsection{Data analysis in stage 2}
\vspace{-0.2cm} 

The quantitative data were compared with the numbers in stage 1. The interview results from the industry were text encoded with: status, challenges, possible solutions, and the feasibility of S-Scrum.

\vspace{-0.2cm}   
\subsubsection{Results in stage 2 -  RQ 4: What are the effects of the optimized S-Scrum on safety and agility?}
\vspace{-0.2cm}

As shown in Fig. 3, most of the evaluated agility aspects sustained a good level of satisfaction with little variance. However, the ``technical design" is slightly reduced. Due to the new role, the collaborative part of design between safety work and development work fell on the internal safety expert. The personal capability is becoming important. To improve the technical design, cooperation shall increase between the external safety expert and the development team.  \par  

\begin{figure*}[!h]
\includegraphics[width=1.0\textwidth]{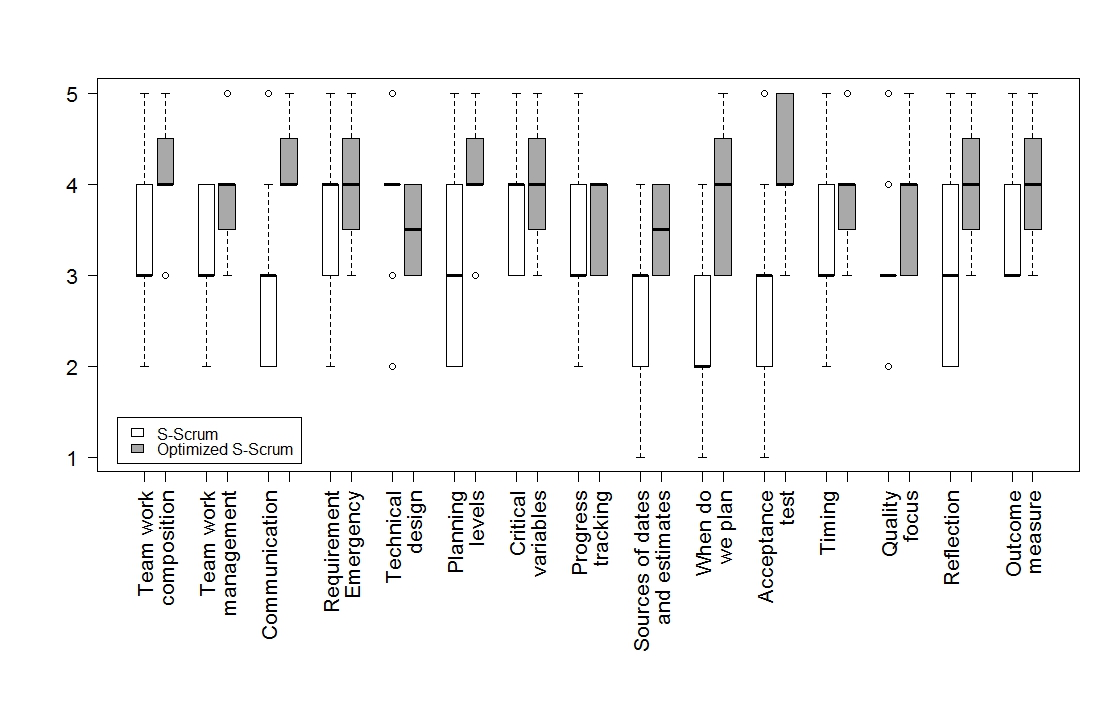}
\caption{Boxplots for agility comparison between S-Scrum and optimized S-Scrum (From ``1" to ``5" means less agile (``negative") to very agile (``positive"))}
\end{figure*}
 
Regarding the safety of optimized S-Scrum, as we can see in Fig. 4, safety aspects improved (M16.1, M16.2, M16.3, M17.1, M17.2). We also rejected few safety requirements (M17.3): 1 (sprint 6), 1 (sprint 7), 0 (sprint 10), 2 (sprint 11). We can conclude that, in general, the optimized S-Scrum has better safety assurance capabilities. However, there are still some abnormal values in sprint 7. The number of safety requirements, the number of safety requirements traceable to hazards and the number of accepted safety requirements in sprint 7 are more than in sprint 10. This may be traced back to the fitting-in phase of the optimized S-Scrum. Since the training of STPA for the internal safety expert, we finished STPA in sprint 10 only once. In sprint 6, sprint 7, and sprint 11, we finished STPA twice. After the adaption of the new role, the safety data rose in sprint 11.

\begin{figure*}[!h]
\center
\includegraphics[width=0.8\textwidth]{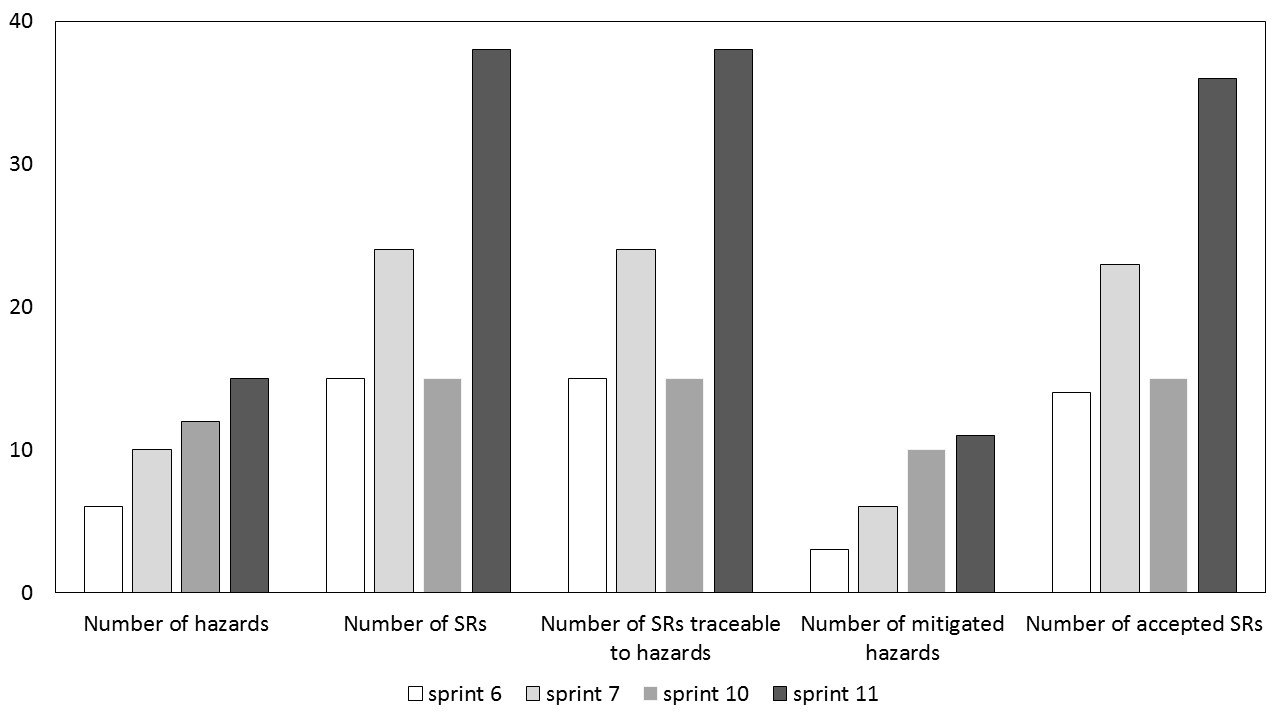}
\caption{Safety data comparison between S-Scrum and optimized S-Scrum (``SRs" - Safety Requirements)}
\end{figure*}

\vspace{-0.4cm}
\subsubsection{Results in stage 2 - Discussion}
\vspace{-0.2cm}
To strength the study further, we discussed our results preliminarily in industry. For \emph{Challenge 1}, the conflict between functional requirements and non-functional requirements seems not obvious. As one interviewee mentioned: \emph{Since we have a relative small amount of non-functional requirements, the priorities are always determined by the product owner together with the discussion with some external experts.} For \emph{Challenge 2}, one interviewee mentioned: \emph{To enhance the communication between the team members and the experts, we have a technical meeting before each sprint planning meeting. The product owner sends the emails to the relevant experts depending on the goals of each sprint. The experts are welcomed to join the daily stand-up meetings.} Thus, the experts have sufficient time to keep up with the development team, while the technical knowledge is deeply discussed in the technical meeting before the sprint planning meeting. The project has also a good knowledge sharing mechanism to support the communication during each sprint. One interviewee mentioned: \emph{We use pair programming, formal guidelines to teach new colleagues, chat clients, and screen sharing. When the team includes experts, the product owner will contact 2-3 colleagues to discuss technical stuff, who will inform other colleagues.} A hierarchical communication mode is preferred for a multi-expert team. For \emph{Challenge 3}, the industrial projects have also mentioned this problem: \emph{Internal user stories are used to record the non-functional requirements. The execution of internal user stories is up to the team.} For \emph{Challenge 4}, the two teams execute a sufficient planning. An interviewee mentioned: \emph{We have a refinement time slot to get all product backlog items approved (each team member has understood) and not so much discussion in the sprint planning meeting.} The team members are beginning the refinement in the present sprint for the user stories in the next sprint. In Scrum, not all requirements have to be at the same level of detail at the same time \cite{rubin2012essential}. The progressive refinement could be further extended for the safety planning and assessment to: (1) avoid a premature development decision from the high-level safety requirements; (2) reserve sufficient time for managing priorities between safety requirements and functional requirements; (3) increase the rework possibilities; (4) enhance the likelihood of using conversation to clarify safety requirements. That could also illustrate the \emph{Challenge 5}. For \emph{Challenge 6}, the refinement phase helps building a pre-understanding of each requirement and reaching a common criterion in the sprint planning meeting.   
The \emph{external expert} is a regular member in industry. An interviewee mentioned: \emph{We prefer some experts with deep knowledge in the team, but the arrangement of an internal expert has to take more issues into account, such as training, responsibility, and even personal development.} An external safety consultant to test the products and delivered trainings and an internal safety initiative \cite{poller2017can} to promote safety practices across groups in industry could be align with our internal and external safety expert. \emph{Safety culture} in industry is enhanced either by setting the regulations or by the established organization structure and activities. An \emph{agile safety plan} is also required from some standards. They draw the safety plan either in the technical meeting or in parallel with the refinement. The technical meeting suggested in industry could also be considered as an \emph{extra (weekly) safety meeting}. The \emph{pre-planning meeting} seems to be a suitable form for realizing progressive refinement in industry. This alignment motivates more combinations between our optimizations and existing industrial practices. All the requirements and \emph{acceptance criteria} are retrieved by \emph{brainstorming}. An effective communication plays a vital role in executing acceptance testing.

\vspace{-0.2cm}
\section{Threats to validity}

\textbf{Construct validity:} The first threat to construct validity is the general data analysis framework. To apply Scrum for safety-critical systems, we focus primarily on safety aspect and agility aspect in our exploratory study. In terms of agility, we referred to an official agility comparative survey \cite{williams2010driving} for ensuring the coverage of measurement. In terms of safety, S-Scrum was extended from Safe Scrum, which was originally developed in accordance with the general functional safety standard IEC 61508. Thus, the validation regarding to the consistency with IEC 61508 has not been included in the framework. Furthermore, in S-Scrum we mainly integrate STPA. We aim to validate the enhanced safety concerning the integrated safety analysis technique. Thus, the safety assurance technique's capability and the deliverable products' safety are set as two relevant goals. Yet, the goals and metrics seem not enough and the validation framework is possible to be extended. The second threat to construct validity is the validation periods for S-Scrum and optimized S-Scrum are shorter than our expectations. We executed the normal Scrum in the first five sprints to strengthen students' background knowledge of agile techniques and prepare the detailed organization structure, which took us a lot of time. \newline 
\textbf{Internal validity:} 
The first threat to internal validity is the arrangement of team roles. One of the authors acted as the product owner and the safety expert concurrently in sprint 6 and sprint 7. To avoid this threat in alignment with the optimizations in sprint 10 and sprint 11, the product owner acted further as an external safety expert. An internal safety expert has been arranged in the development team. The second threat to internal validity exists in the qualitative data from the semi-structured interviews. The interviews have been performed by one of the authors together with the audio record. The language we used has also partial German. To avoid subjective and language bias, the audio recording has been transcribed independently by two researchers (one is a native German speaker) and compared to formulate a final result.\newline 
\textbf{External validity:} A student project is different from an industrial project. However, H{\"o}st et al. \cite{host2000using}, Tichy, Kitchenham et al. \cite{tichy2000hints} proposed that students could be acceptable. To consider this debatable issue, we mainly referred to an empirical study conducted by Falessi in 2017 \cite{ese}. 16 statements are provided by 65 empirical researchers. They mentioned: \emph{Conducting experiments with professionals as a first step should not be encouraged unless high sample sizes are guaranteed or performing replicas is cheap.} In our research, there exists few industrial projects for developing safety-critical systems fully adopted a Scrum development process according to the preliminary research \cite{theo}. S-Scrum was also proposed in 2016 as a high-level process model. In addition, the long learning cycles and a new technology are two hesitations for using professionals. STPA was developed in 2012. In industry, there is still a lack of experts. Thus, we believe that in our research area, a student project is a relative suitable way to aggregate contributions. Even though, the generalizability is considered critical. \newline 
\textbf{Reliability:} The student project is a suitable way for a first validation. Yet, the results from the students are limited by their personal experience. Besides, the ``grading power" of the researchers may influence the results. We separated our research work from the final examination of the product to mitigate this threat.       
\vspace{-0.4cm}
\section{Conclusion} 
\vspace{-0.3cm} 

The main benefit of our research is that it provides a first empirical and practical insight into applying Scrum for safety-critical systems with the integration of STPA. Moreover, the presented challenges existing in priority management, communication, time pressure on determining safety requirements, safety planning, safety requirements' acceptance criteria and solutions including the split of the safety expert, pre-planning meeting, regular safety meeting, improved safety epics, STPA-based safety stories and an agile safety plan could arouse interest in practitioners and show future research directions. The effects on safety and agility aspects indicate the feasibility to align STPA with a Scrum development process. The discussion in industry motivates the further step of transmitting the optimized S-Scrum from the academic environment towards industry environment. However, the execution of S-Scrum and optimized S-Scrum was in a specific context. We can rely our improvements on an academical project only. The generalization in industry of the optimizations remains subject to future work. Finally, regarding safety and security in agile development in today's cyber-physical systems, even though special attention has to be paid to the respective norms and standards, problems' exploration in practice seems also necessary.    \par

\section{Acknowledgements.} 
\vspace{-0.3cm} 
 \footnotesize 
 We want to thank Dr. A. Nguyen-Duc for proof reading and his valuable suggestions. We are grateful to all participants involved during the case study. Finally, we want to thank all the feedback on previous versions. The first author is supported by the LGFG (Stipendien nach dem Landesgraduiertenf{\"o}rdergesetz).


\vspace{-0.3cm}
\begin{thebibliography}{4}
\vspace{-0.3cm}
\footnotesize

\bibitem{iec61508}IEC61508, Functional safety of electrical/electronic/programmable electronic safety-related systems. International Electrotechnical Commission, 2010.

\bibitem{turk2014limitations}Turk, D., France, R., and Rumpe, B.: Limitations of agile software processes. arXiv preprint arXiv:1409.6600 (2014).

\bibitem{staalhane2012application}St{\aa}lhane, T., Myklebust, T., and Hanssen, G. K.: The application of Safe Scrum to IEC 61508 certifiable software. 11th International Probabilistic Safety Assessment and Management Conference and the Annual European Safety and Reliability Conference. 2012.

\bibitem{staalhane2013scrum}St{\aa}lhane, T., Vikash K., and Myklebust, T.: Scrum and IEC 60880. Enlarged Halden Reactor Project meeting, Storefjell, Norway. 2013.

\bibitem{staalhanesafety}St{\aa}lhane, T.: Safety standards and Scrum – A synopsis of three standards.

\bibitem{hanssen2016quality}Hanssen, G.K., et al. Quality Assurance in Scrum Applied to Safety Critical Software. International Conference on Agile Software Development. Springer International Publishing, 2016.

\bibitem{leveson2011engineering}Leveson, N.: Engineering a safer world: Systems thinking applied to safety. MIT press, 2011.

\bibitem{ge2010iterative}Ge, X., Richard F.P., John, A. M.: An iterative approach for development of safety-critical software and safety arguments. AGILE Conference. IEEE, 2010.

\bibitem{vuori2011agile}Vuori, M.: Agile development of safety-critical software. Tampere University of Technology 14 (2011).

\bibitem{diebold2015practitioners}Diebold, P., et al.: What do practitioners vary in using scrum?. International Conference on Agile Software Development. Springer International Publishing, 2015.

\bibitem{cho2010exploratory}Cho, J.J.: An exploratory study on issues and challenges of agile software development with \protect{S}crum. All Graduate Theses and Dissertations (2010): 599.

\bibitem{williams2010driving}Williams, L., Kenny R., and Mike, C.: Driving process improvement via comparative agility assessment. AGILE Conference, 2010. IEEE, 2010.

\bibitem{cruickshank2009validation}Cruickshank, K.J., James, B.M., and Man-Tak, S.: A validation metrics framework for safety-critical software-intensive Systems. System of Systems Engineering, 2009. SoSE 2009. IEEE International Conference. IEEE, 2009.

\bibitem{kelly2004goal}Kelly, T., and Rob, W.: The goal structuring notation – a safety argument notation. Proceedings of the dependable systems and networks 2004 workshop on assurance cases. Citeseer, 2004.

\bibitem{basili1992software}Basili, V.R.: Software modeling and measurement: the goal/question/metric paradigm. 1992.

\bibitem{wang2016toward}Wang, Y., and Wagner, S.: Toward Integrating a System Theoretic Safety Analysis in an Agile Development Process. Software Engineering. 2016.

\bibitem{runeson2009guidelines}Runeson, P, H{\"o}st, M.: Guidelines for conducting and reporting case study research in software engineering. Empirical software engineering 14.2 (2009): 131.

\bibitem{yin2013case}Yin, R. K.: Case study research: Design and methods. Sage publications, 2013.

\bibitem{strauss1997grounded}Strauss, A., and Corbin, J.M.: Grounded theory in practice. Sage, 1997.

\bibitem{poller2017can} Poller, A., Kocksch, L., T{\"u}rpe, S., Epp, F.A., and Kinder-Kurlanda, K.: Can security become a routine?: a study of organizational change in an agile software development group. Proceedings of the 2017 ACM Conference on Computer Supported Cooperative Work and Social Computing. ACM, 2017.

\bibitem{myklebust2016agile}Myklebust, T., St{\aa}lhane, T., and Lyngby, N.: The agile safety plan. PSAM13 (2016).

\bibitem{myklebust2016safety}Myklebust, T., St{\aa}lhane, T.: Safety stories – a new concept in agile development. Fast Abstracts at International Conference on Computer Safety, Reliability, and Security (SAFECOMP 2016). 2016.

\bibitem{garg2015cucumber}Garg, S.: Cucumber cookbook. Packt Publishing Ltd, 2015.

\bibitem{rubin2012essential}Rubin, K.S.: Essential scrum: A practical guide to the most popular Agile process. Addison-Wesley, 2012.

\bibitem{Moe:2010:TMU:1752257.1752480}Moe, N.B., Torgeir D., and Tore D.: A teamwork model for understanding an agile team: A case study of a Scrum project. Information and Software Technology 52.5 (2010): 480-491.

\bibitem{begel2007usage}Begel, A., and Nachiappan N.: Usage and perceptions of agile software development in an industrial context: An exploratory study. Empirical Software Engineering and Measurement, 2007. ESEM 2007. First International Symposium on. IEEE, 2007.

\bibitem{turner2002agile}Turner, R., and Apurva J.: Agile meets CMMI: Culture clash or common cause?. Conference on Extreme Programming and Agile Methods. Springer Berlin Heidelberg, 2002.

\bibitem{mccaffery2008ahaa}McCaffery, F., Pikkarainen, M. and Richardson, I.: Ahaa--agile, hybrid assessment method for automotive, safety critical smes. Proceedings of the 30th international conference on Software engineering. ACM, 2008.

\bibitem{moe2012challenges}Moe, N.B., Ayb{\"u}ke A., and Dyb{\aa}, T.: Challenges of shared decision-making: A multiple case study of agile software development. Information and Software Technology 54.8 (2012): 853-865.


\bibitem{host2000using}H{\"o}st, M., Bj{\"o}rn R., and Wohlin, C.: Using students as subjects — a comparative study of students and professionals in lead-time impact assessment. Empirical Software Engineering 5.3 (2000): 201-214.

\bibitem{tichy2000hints}Tichy, W.F.: Hints for reviewing empirical work in software engineering. Empirical Software Engineering 5.4 (2000): 309-312.

\bibitem{kitchenham2002preliminary}Kitchenham, Barbara A., et al.: Preliminary guidelines for empirical research in software engineering. IEEE Transactions on software engineering 28.8 (2002): 721-734.

\bibitem{wang}Wang, Y, and Wagner, S.: Towards applying a safety analysis and verification method based on STPA to agile software development. IEEE/ACM International Workshop on Continuous Software Evolution and Delivery (CSED). IEEE, 2016.

\bibitem{wangdc}Wang, Y, Bogicevic, I. and Wagner, S.: A study of safety documentation in a Scrum development process. Proceedings of the XP2017 Scientific Workshops. ACM, 2017.

\bibitem{theo}Theocharis, G., Kuhrmann, M., M{\"u}nch, J., and Diebold, P.: Is water-scrum-fall reality? on the use of agile and traditional development practices. In International Conference on Product-Focused Software Process Improvement (pp. 149-166). Springer International Publishing.

\bibitem{ese}Falessi, D., Juristo, N., Wohlin, C., Turhan, B., M{\"u}nch, J., Jedlitschka, A., and Oivo, M.: Empirical software engineering experts on the use of students and professionals in experiments. In Journal of Empirical Software Engineering. Springer, 2017.



\end{thebibliography}
\end{document}